\documentclass{iopjournal}

\usepackage[square]{natbib}
\usepackage{lineno}
\usepackage{soul}
\usepackage{xcolor}
\usepackage{hyperref}
\usepackage[normalem]{ulem}
\usepackage{graphicx}
\usepackage{siunitx} 
\usepackage{booktabs}
\usepackage{lmodern}
\usepackage[T1]{fontenc}
\usepackage{ragged2e}

\usepackage{fancyhdr}
\begin{document}

\justifying

\title{Fast silicon detectors as proton therapy beam monitors}

\author{
A. Bellora$^1$\orcid{0000-0002-2753-5473}, 
L. Grzanka$^{1,2}$\footnote{Corresponding author: \texttt{grzanka@agh.edu.pl}}\orcid{0000-0002-3599-854X}, 
R. McNulty$^3$\orcid{0000-0001-7144-0175}, 
N. Minafra$^4$\orcid{0000-0003-4002-1888}, 
K. Misan$^1$\orcid{0009-0007-2416-042X}, 
M. Nessel$^1$\orcid{0009-0007-6981-1396}, 
T. Nowak$^2$\orcid{0000-0002-5360-1008}, 
P. Rzeznik$^1$\orcid{0009-0006-9868-7307}, 
J. Swakon$^2$\orcid{0000-0003-4002-1888} and
T. Szollosova$^5$\orcid{0009-0001-8072-3748}
}

\affil{$^1$AGH University of Krakow, Poland}
\affil{$^2$Institute of Nuclear Physics, Polish Academy of Sciences, Krakow, Poland}
\affil{$^3$School of Physics, University College Dublin, Belfield, Dublin 4, Ireland}
\affil{$^4$Department of Physics \& Astronomy, University of Kansas, Lawrence, KS 66045, USA}
\affil{$^5$Faculty of Nuclear Sciences and Physical Engineering, Czech Technical University in Prague}

\email{grzanka@agh.edu.pl}

\begin{abstract}
Low Gain Avalanche Detectors (LGADs) are fast silicon sensors with potential for high-resolution proton beam monitoring. In this work, we investigated their response in a \SI{60}{\mega\electronvolt} proton beam from an AIC-144 cyclotron.
Several features of the pulse structure of the cyclotron were observed for the first time.
The detector and acquisition system tested was shown to provide accurate flux measurements up to an instantaneous dose rate of about \SI{7.5}{\gray\per\second}.
Above this, there is an efficiency loss due to multiple protons crossing the detector, which could be addressed through a redesign of the sensors and electronics.
LGADs therefore offer a viable alternative for proton therapy beam diagnostics.

\end{abstract}


\section{Introduction}
\label{sec:intro}

Accurate beam monitoring represents a crucial component in proton therapy, ensuring precise dose delivery and optimal treatment outcomes. Modern proton therapy facilities typically employ cyclotrons or synchrotrons as particle accelerators, each introducing distinct temporal beam characteristics that must be carefully monitored \citep{paganetti2025proton}. Cyclotrons operate with a radio frequency (RF) acceleration system typically in the range of \SIrange{50}{100}{\mega\hertz}, creating a bunch structure with spacing in the order of \SIrange{10}{20}{\nano\second}, while synchrotrons deliver protons in longer spills lasting several seconds \citep{schippers2009beam}.

The implementation of pencil beam scanning (PBS) technology has highlighted the need for high spatial and temporal resolution devices to monitor the hadron beam with intensity over several orders of magnitude (wide dynamic range)~\citep{li2012beyond, rydygier2018radiotherapy} to ensure precise dose calculations.
Traditional beam monitoring devices, such as ionization chambers coupled with electrometers, while robust and well-established, face inherent temporal limitations. The combined response time of an ionisation chamber and electrometer system typically exceeds \SI{500}{\micro\second} \citep{newhauser2009international}. A detector system with a few microseconds response would therefore significantly outperform any conventional ionisation chamber setup in temporal resolution. Silicon-based detection systems, particularly advanced designs that incorporate timing optimisation, present a promising solution to these challenges. 

The rapid signal formation characteristics of modern silicon sensors (few nanoseconds) make them suitable for operation in particle counting mode even with bunch spacing as short as \SI{10}{\nano\second}, typical for \SI{100}{\mega\hertz} RF systems. Beam monitoring instrumentation  used in proton therapy facilities can be exposed to high radiation fluence, with annual proton fluences extending approximately \SI{e15}{\per\square\centi\meter}. Ensuring stable operation under these conditions requires radiation-hard sensors, where any accumulated damage does not significantly distort particle count accuracy and beam monitoring performance.

Low Gain Avalanche Detectors (LGADs) \citep{pellegrini2014technology} 
are starting to find applicability in medical physics~\citep{MARTIVILLARREAL2023167622,monaco2023performance,lgad_heller} as they
offer a particularly promising solution for beam monitoring applications, with subnanosecond timing resolution,  small pixel sizes ($<$0.5 mm), and a wide dynamic range, allowing real-time monitoring of high-intensity regions and low-dose tails.
LGADs are characterised by a shallow \SI{\sim 1}{\micro\meter} gain zone and a high electric field configuration, enabling rapid signal development while maintaining low gain characteristics. This design requires specialised low-noise electronics for signal readout but achieves remarkable timing performance. These detectors can produce signals with widths below \SI{2}{\nano\second} and measure particle passage times with resolutions approaching \SI{10}{\pico\second}. Recent developments in LGAD technology, driven by the demands of detecting separate proton-proton interactions within colliding bunches at the Large Hadron Collider (LHC), have led to significant improvements in both performance and radiation hardness \citep{Albrow:2014lrm}.

For practical implementation in clinical proton therapy, the maximum measurable particle flux must be considered within the constraints of detector timing capabilities. Consider a detector with an active area of \SI{1}{\square\milli\meter} operating with a cyclotron RF frequency of \SI{100}{\mega\hertz}, corresponding to \SI{10}{\nano\second} between bunches. To maintain a pileup probability below \SI{10}{\percent}, ensuring accurate particle counting, the detector must process no more than one particle per ten RF periods on average. This translates to a maximum instantaneous particle flux during the beam spill of \SI{10}{\mega\hertz} per \si{\square\milli\meter}, or \SI{e9}{\per\square\centi\meter\per\second}. For protons with a kinetic energy of \SI{120}{\mega\electronvolt}, this instantaneous flux corresponds to a dose rate (in water) of approximately \SI{1}{\gray\per\second}. Such flux capabilities are well-matched to typical therapeutic proton beam intensities, (which are on the order of \SI{1}{\gray\per\minute} or even \SI{0.5}{\gray\per\second} for hyperfractionated treatment of eye tumours) while maintaining precise temporal resolution.

In previous work, we examined the response of a LGAD in an electron beam of an Elekta linac~\citep{Isidori:2021cwt} and showed its potential as a beam monitor and its ability to resolve the linac temporal structure.
Improvements to the front-end amplifier have shortened the signal times by about an order of magnitude allowing single charged particles in higher fluence environments to be distinguished.

In this study, we evaluate LGADs as beam monitors in a proton beam produced by the AIC-144 cyclotron at the Institute of Nuclear Physics, Kraków. The 58\si{\mega\electronvolt} cyclotron, now dedicated to research and previously used in the treatment of ocular melanoma, was operated in a standard clinical environment. The LGAD response was characterised using large-area air ionisation chambers and electrometers. The low duty cycle of the cyclotron allowed exploration of the behaviour of the LGAD at instantaneous particle rates of 3 to 360 \si{\mega\hertz} and pileup factors from 0.1 to 13, while maintaining mean dose rates below 1\si{\gray\per\second}. This allowed us to access pileup regimes higher than those previously studied \citep{monaco2023performance}.

The fast timing capabilities of LGADs enabled the observation of cyclotron beam characteristics that were previously unresolved. In addition it provided new insights into beam dynamics and detector performance.

The paper is organised as follows.
The beamline and LGAD detectors are described in Section~\ref{sec:apparatus} together with the methods developed to identify LGAD signals, investigate the pulse structure of the cyclotron beam, and monitor the beam delivery. 
In Section~\ref{sec:results} we present our measurements. 
The results and beam monitoring methods are discussed and compared in Section~\ref{sec:discussion}. 
Our conclusions are presented in Section~\ref{sec:conclusions}

\section {Apparatus and methodology}
\label{sec:apparatus}

\subsection{Proton beam}
\label{sec:cyclotron}

\begin{figure}[!ht]
\centerline{\includegraphics[scale=0.8]{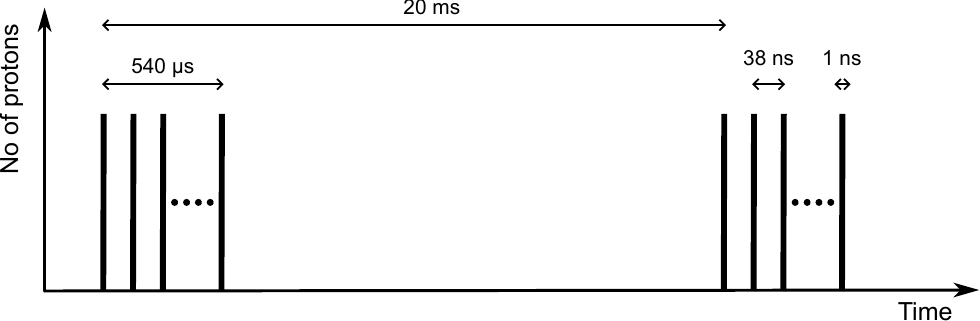}}
\caption{
Schematic profile of the nominal time structure of the cyclotron pulses, showing two consecutive macropulses, each of \SI{540}{\micro\second} duration. Each macropulse consists of \SI{1}{\nano\second} micropulses separated by about \SI{38}{\nano\second}.
}
\label{fig:sketch}
\end{figure}

The AIC-144 isochronous cyclotron at the Institute of Nuclear Physics in Krakow accelerates protons to a kinetic energy of \SI{58}{\mega\electronvolt} \citep{swakon2010facility}. LGAD sensors were irradiated on the fixed horizontal passive beamline, previously used for radiotherapy of uveal melanoma tumours. The beam was laterally spread on a single \SI{25}{\micro\meter} tantalum foil located \SI{11}{\meter} upstream of the isocenter. A \SI{40}{\milli\meter} brass collimator was used to produce a flat, homogeneous beam that covered the entire sensor and a large part of the amplifier board.

The mean beam intensity can be adjusted from \SI{0.01}{\gray\per\second} (\SI{5e6}{protons\per\centi\meter\squared\per\second}) to \SI{10}{\gray\per\second} (\SI{5e9}{protons\per\centi\meter\squared\per\second}). The nominal temporal structure of the beam is illustrated in Fig.~\ref{fig:sketch}. Every \SI{20}{\milli\second} (corresponding to \SI{50}{\hertz}), a macropulse (short spill) is generated, lasting approximately \SI{540}{\micro\second}. Within each macropulse, micropulses (bunches) of about \SI{1}{\nano\second} in duration are produced at a radio frequency (RF) of \SI{26.26}{\mega\hertz}. The beam current is measured every \SI{0.1}{\second} using ion chambers positioned at various distances from the isocentre.

\subsection{Beam monitoring}
\label{sec:beam_monitoring}

The instantaneous flux, $\Phi_{\text{inst}}$, is defined as the number of protons, $N_{\text{inst}}$, per macropulse duration, $T_{\text{macrop}}$, normalized by the detector area, $S$:
\begin{equation}
    \Phi_{\text{inst}} = \frac{N_{\text{inst}}}{S \cdot T_{\text{macrop}}}.
    \label{eq:phi_inst}
\end{equation}
For the studied system, $T_{\text{macrop}} \approx \SI{540}{\micro\second}$. Note that any uncertainty in the determination of $T_{\text{macrop}}$ propagates directly into $\Phi_{\text{inst}}$. The instantaneous flux neglects beam-intensity fluctuations on smaller timescales (i.e. over a few \SI{}{\micro\second}).

 The mean flux, $\Phi_{\text{mean}}$, accounts for the cyclotron duty cycle and is defined as the total number of protons, $N_{\text{mean}}$, in an arbitrary measuring period, $T_{\text{arb}}$, containing multiple macro-pulses, normalized by the detector area:
\begin{equation}
    \Phi_{\text{mean}} = \frac{N_{\text{mean}}}{S \cdot T_{\text{arb}}}.
    \label{eq:phi_mean}
\end{equation}
The mean quantity is not dependent on the exact macro-pulse length and can be compared with readings of monitoring ionisation chambers. 
In a similar way to how the mean and instantaneous particle fluxes are defined, a mean and instantaneous dose rate can be obtained. 

The dose rate, $\dot{D}$, is related to the proton flux, $\Phi$, by 
\begin{equation}
    \label{eq:dose_fluence}
    \dot{D} \cdot \rho_{\text{w}} = \Phi \cdot S_{\text{w}},
\end{equation}
where
$ \rho_{\text{w}}$  is the density of water and $ S_{\text{w}} $ is the electronic stopping power of protons in water.  
The kinetic energy of protons at the isocentre was assumed to be \SI{58.4}{\mega\electronvolt}, which corresponds to 
\( S_{\text{w}} = \SI{11.22}{\mega\electronvolt\per\centi\meter} \), and
was calculated using the SRIM programme \citep{ziegler2010srim}.
In our study, equation \ref{eq:dose_fluence} was used to relate the mean and instantaneous dose rates to the proton flux.

The AIC-144 cyclotron has a low duty cycle of \SI{2.7}{\percent}, operating for \SI{540}{\micro\second} every \SI{20}{\milli\second}. Consequently, a mean dose rate of \SI{6}{\gray\per\minute} corresponds to a mean flux of \SI{5e7}{protons\per\centi\metre\squared\per\second} and a mean particle rate of \SI{1.1}{\mega\hertz} through a \SI{2.2}{\milli\metre\squared} detector. This corresponds to an instantaneous dose rate (within a macropulse) of \SI{220}{\gray\per\minute} (\SI{3.7}{\gray\per\second}) and to an instantaneous flux of \SI{2e9}{protons\per\centi\metre\squared\per\second} and a particle rate of \SI{44}{\mega\hertz}.

\setlength{\tabcolsep}{4pt}
\begin{table}[ht]
\centering
\resizebox{\textwidth}{!}{
\begin{tabular}{@{}lS[round-mode = places, round-precision=2]S[exponent-mode = threshold, exponent-thresholds=-4:4,round-mode = places, round-precision=3]S[round-mode = places, round-precision=2]S[exponent-mode = threshold, exponent-thresholds=-3:3,round-mode = places, round-precision=2]S[exponent-mode = threshold, exponent-thresholds=-3:3, round-mode=places, round-precision=2]@{}}
\toprule

dataset & \multicolumn{1}{c}{mean flux} & \multicolumn{1}{c}{mean dose rate} & \multicolumn{1}{c}{typical inst. flux} & \multicolumn{1}{c}{typical inst.} & \multicolumn{1}{c}{expected} \\

& \multicolumn{1}{c}{$(\Phi^{\text{IC}}_{\text{mean}}$)} & \multicolumn{1}{c}{} & \multicolumn{1}{c}{$(\Phi^{\text{IC}}_{\text{inst}})$} & \multicolumn{1}{c}{dose rate} & \multicolumn{1}{c}{prot./micropulse} \\

& \multicolumn{1}{c}{[\si{\per\centi\meter\squared\per\second}]}  & \multicolumn{1}{c}{[\si{\gray\per\second}]}             & \multicolumn{1}{c}{[\si{\per\centi\meter\squared\per\second}]}  & \multicolumn{1}{c}{[\si{\gray\per\second}]} & \\
\midrule
0.5nA & 3.606e+06 & 6.482e-03 & 1.336e+08 & 2.401e-01 & 1.119e-01\\
1nA & 7.616e+06 & 1.369e-02 & 2.821e+08 & 5.071e-01 & 2.363e-01\\
2nA & 1.609e+07 & 2.893e-02 & 5.960e+08 & 1.071e+00 & 4.993e-01\\
4nA & 2.836e+07 & 5.098e-02 & 1.050e+09 & 1.888e+00 & 8.801e-01\\
8nA & 5.864e+07 & 1.054e-01 & 2.172e+09 & 3.904e+00 & 1.819e+00\\
16nA & 1.176e+08 & 2.113e-01 & 4.354e+09 & 7.827e+00 & 3.648e+00\\
32nA & 2.140e+08 & 3.847e-01 & 7.927e+09 & 1.425e+01 & 6.641e+00\\
64nA & 4.418e+08 & 7.942e-01 & 1.636e+10 & 2.942e+01 & 1.371e+01\\
\bottomrule
\end{tabular}
}

\caption{Summary of the beam intensities used in the exploration of the LGAD sensor.
The first column labels the dataset using
the nominal beam current of the cyclotron, which is correlated with the current measured by the ionisation chambers. The second and third columns give the mean proton flux and corresponding dose rate derived from the ionisation chamber measurements. 
The fourth and fifth columns give typical instantaneous fluxes and the corresponding dose rate. 
The sixth column presents the expected number of protons per micropulse.
}
\label{tab:design}
\end{table}

During measurements with the LGAD detectors, two parallel plate ionisation chambers were positioned on the beamline. The dimensions of these chambers exceeded the beam envelope, ensuring that the full beam current was recorded. Before the experiment, the charge \( Q_{\text{cal,em}} \) measured by the beamline ionisation chambers was calibrated against the dose \( D_{\text{cal,iso}} \) at the isocenter, which was measured using a Markus ionization chamber. The Markus chamber, due to its smaller size, sampled only a portion of the beam within the treatment field, while the beamline chambers integrated the total beam current.  

The mean proton flux at the isocenter, \( \Phi^{\text{IC}}_{\text{mean}} \), was determined using the current recorded by the ionization chamber,
\( I_{\text{em}}\), from the following relation:  
\begin{equation}
    \Phi^{\text{IC}}_{\text{mean}} = I_{\text{em}} \cdot \frac{D_{\text{cal,iso}}}{Q_{\text{cal,em}}} \cdot \frac{\rho_{\text{w}}}{S_{\text{w}}}.
\end{equation}

The ionisation chamber readings were recorded at \SI{100}{\milli\second} intervals (\(T_{\text{IC}}\)), each covering \(n=5\) macropulses. Assuming negligible beam fluctuations among these macropulses, the instantaneous flux \(\Phi^{\text{IC}}_{\text{inst}}\) can be expressed as  
\begin{equation}
    \Phi^{\text{IC}}_{\text{inst}} = \frac{T_\text{IC}}{n T_{\text{macrop}}} \Phi^{\text{IC}}_{\text{mean}}.
    \label{eq:flux_ic_inst}
\end{equation}  
For the system studied, the scaling factor between the mean and instantaneous fluxes 
\linebreak$\Phi^{\text{IC}}_{\text{inst}}
/
\Phi^{\text{IC}}_{\text{mean}}=
T_\text{IC} / (n T_{\text{macrop}})$
~was $37.04 \pm 0.01$.

The expected number of protons traversing the sensor during each micropulse can be determined as 
\begin{equation}
    N^{\text{exp}}_{\text{micro}}=
    \Phi^{\text{IC}}_{\text{inst}}
    S
    \frac{T_{\text{RF}}}{T_{\text{macro}}},
    \label{eq:exp_prot_per_microp}
\end{equation}
where $S$ is the sensor area (described in Sec.~\ref{sec:sensors}), $T_{\text{RF}}$ is the RF period of the cyclotron (computed in Sec.~\ref{sec:res_rf}), and $T_{\text{macro}}$ is the nominal macropulse duration (\SI{540}{\micro\second}). 

The proton beam intensities, used in the evaluation of the performance of the LGAD, are summarised in Table \ref{tab:design}. 
In our studies, the instantaneous flux from the LGAD measurements, $\Phi^{\text{LGAD}}_{\text{inst}}$,
obtained using Eq.~\ref{eq:phi_inst},  is compared to \(\Phi^{\text{IC}}_{\text{inst}}\) calculated using the ionisation chamber measurements (Eq.~\ref{eq:flux_ic_inst}). 
The beam uniformity within the sensor dimensions was verified through a transverse scan, performed using standard QA beam quality methods, and demonstrated a consistent detector response across the scanned area.

\subsection{LGAD sensors}
\label{sec:sensors}

\begin{figure}[!ht]
\centerline{

\includegraphics[width=0.48\textwidth]{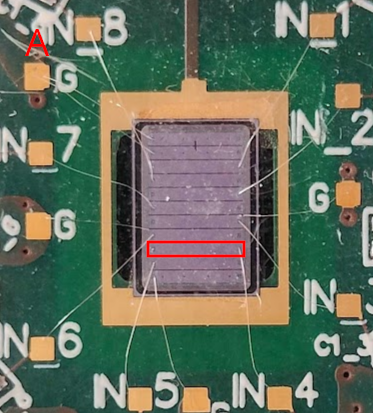}
\includegraphics[width=0.48\textwidth]{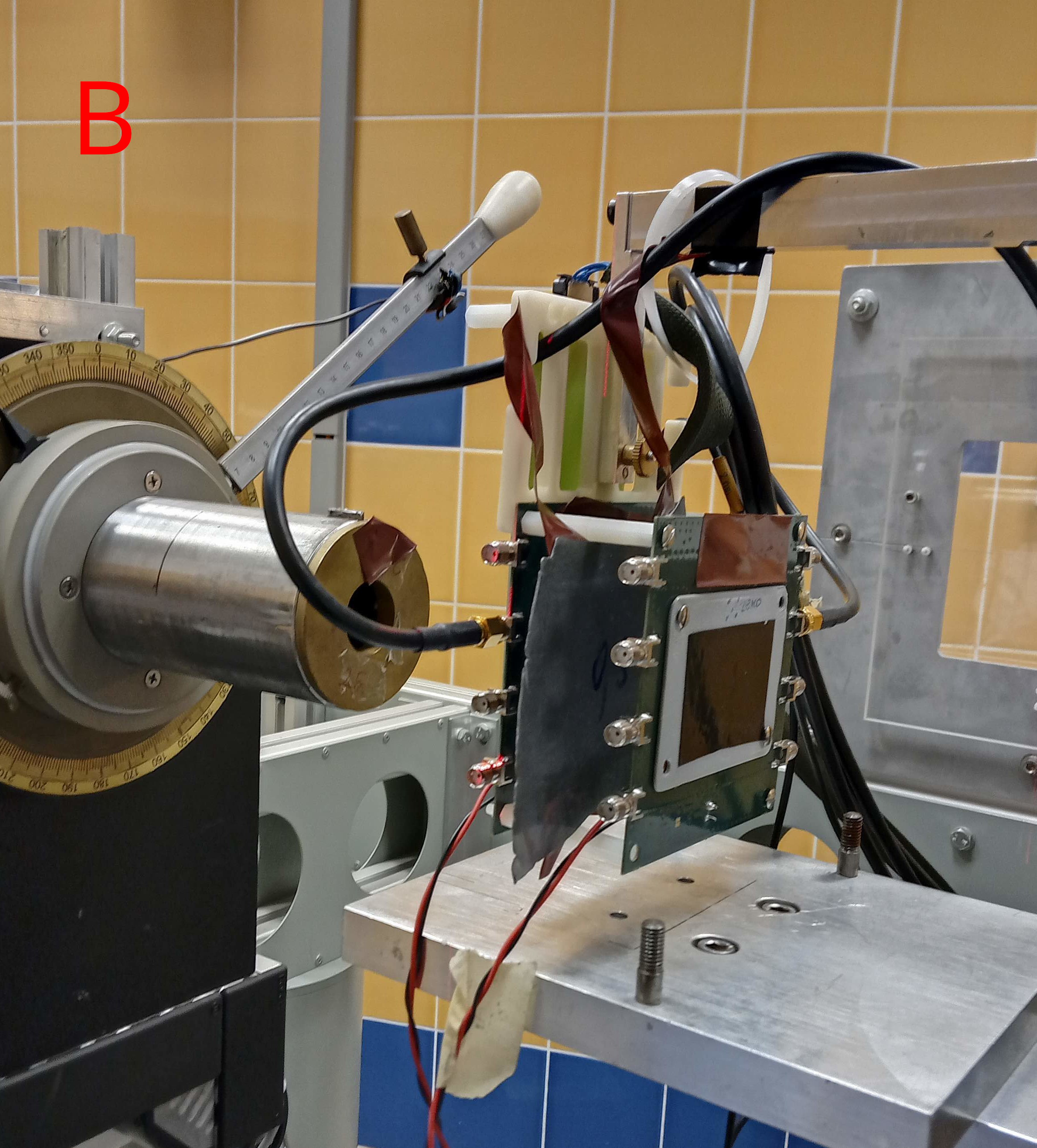}
}
\caption{
Panel A (left): The LGAD sensor mounted on the readout board. The strip used in this study is highlighted in red.
Panel B (right): Two sensors mounted one behind the other in the cyclotron beamline. The first board is positioned in the isocenter. }
\label{fig:lgad}
\end{figure}

The LGAD sensors used in this study were originally developed at the Fondazione Bruno Kessler (FBK, Trento) for the MoVe-IT\footnote{\url{https://www.tifpa.infn.it/projects/move-it/}} project (“type T” from the MoVeIT-2020/UFSD3 production batch\citep{shakarami2021development,abujami2024development}). Each sensor has 11 strips, each with an active area measuring \SI{4.0}{\milli\metre} $\times$ \SI{0.55}{\milli\metre} and an active thickness of \SI{55}{\micro\metre}, as shown in 
Fig.~\ref{fig:lgad}(left).  
Eight strips were bonded to the front-end board, while three others (two upper and one lower) and the guard ring were grounded. 
Two boards were installed in the beamline, separated by \SI{4}{\centi\metre} (as shown in Fig.~\ref{fig:lgad} (right)).
Three channels on the first board were read out by the data-acquisition system, and one on the second board.
The sensor was biased at \SI{200}{\volt}, its nominal operating point, providing a gain of approximately 20.

LGAD sensors are commonly used for precise timing measurements of minimum ionising particles (MIPs) and their front-end amplifiers are usually designed for an optimal time precision on the leading edge of the signal. 
However, the \SI{60}{\MeV} protons employed here deposit significantly more energy, around five times more than MIPS, and the instantaneous fluxes are greater than in typical particle physics applications.
Consequently, the amplifier was optimised to generate a fast signal, i.e.\ a faster rise time and a shorter recovery time, at the expense of a poorer signal-to-noise ratio, in order to improve the sensor's high-fluence performance.

Silicon detectors are known to be very efficient devices, with detection efficiencies above 98\%. However, phenomena such as charge sharing, cross-talk, and pileup can reduce the counting efficiency.

Charge sharing consists in the collection of electron-hole pairs, produced by a single particle traversing the detector, being collected by multiple electrodes. Effectively, this phenomenon decreases the amount of collected charge and potentially decreases the amplitude to values lower than the detection threshold, creating inefficiency. These effects have been found to be negligible in LGADs, due to the high electric field inside the gain layer: in \citep{abujami2024development}, the charge sharing effect for carbon ions is reported to be less than 1\% for the sensor used in this study.

Cross-talk takes place when charge migration towards the readout electrode induces a voltage pulse of opposite sign in neighbouring electrodes. The amplitude of the induced pulse (positive in our case) is correlated with that of the negative signal. 
Whenever an induced pulse overlaps with the one produced by another particle traversing the sensor, some inefficiency may arise. In fact, if the superposition of the two pulses decreases the negative signal below the detection threshold, the particle will not be counted.
In the data collected for this work, no correlation was found between the amplitudes of coincident negative and positive pulses in different detector channels, thus excluding the presence of cross-talk. However, we observed the presence of positive pulses that are synchronous with the micropulse cycle. The source of these pulses has not been fully determined but may be induced by beam particles hitting the transistors of the front-end amplifier, or by electromagnetic interferences picked up by the bonding wires, and typically induce a voltage response similar to that of a dampened oscillator, the negative part of which might be identified as a pulse induced by a particle hitting the sensor. Effects related to these pulses have been reduced by excluding subsequent negative pulses following the first signal within the same micropulse.
However, they can influence particle counting when a large positive pulse is superimposed on a genuine (negative) signal, which was found to be particularly significant at high beam intensities, and is discussed further in Section~\ref{sec:dosimetry_results}.

Pileup occurs when two particles traverse the same detector channel within a time shorter than the signal width of about \SI{1}{\nano\second}, such that two separate signals are not produced. In this case a single singal will be recorded, though usually with a greater amplitude or a broader profile. Since these events are counted as a single particle, they contribute to the inefficiency of particle counting. 
Pileup is unlikely to significantly affect the counting efficiency at low particle fluences. However, at higher fluences, there is a non-negligible probability of two or more particles crossing the detector within the time for the signal width and thus, pileup contributions can have a significant effect.

\subsection{Data acquisition}

The acquisition system recorded the waveforms for subsequent offline analysis. This approach offers a significant advantage for detailed timing analysis, including mitigation of time-walk effects, verification of coincidences, and investigation of overlapping pileup signals. Inspecting the recorded waveforms allows for a precise definition of peak counting criteria and enables a thorough investigation of systematic effects, such as cross-talk between channels and beam-induced electrical effects. Positive peaks, for example, can distort the detection of negative particle signals through cancellation effects. Instead of a dedicated data acquisition system, a LeCroy WavePro 404HD oscilloscope was used. The three channels on the upstream board and the central channel on the downstream board were connected to the oscilloscope, which digitised the signals at \SI{10}{\giga\mathrm{Samples\per\second}} with a 12-bit ADC. All reported results in this study originate from a single strip of the upstream sensor, highlighted in Fig.~\ref{fig:lgad}.

Data were acquired over consecutive \SI{1}{\milli\second} time windows triggered by a \SI{50}{\hertz} cyclotron signal.  Each trigger captured 25 complete macropulses corresponding to \SI{0.5}{\second} of the beam time and stored the data locally on the oscilloscope before transfer to an external SSD.  Data transfer took approximately \SI{7}{\second}.
Data were collected for several minutes at the different beam intensities summarised in Table~\ref{tab:design}, resulting in tens of GBs of waveform data per dataset. Such a long data-capture time combined with high acquisition efficiency allowed us to capture previously unseen features of the time evolution of the beam at a microsecond scale. The currents from the ionisation chambers were recorded every 0.1 s and synchronised with the LGAD waveforms by the timestamps assigned to both datasets.

\begin{figure}[!ht]
\centerline{
\includegraphics[width=16cm]{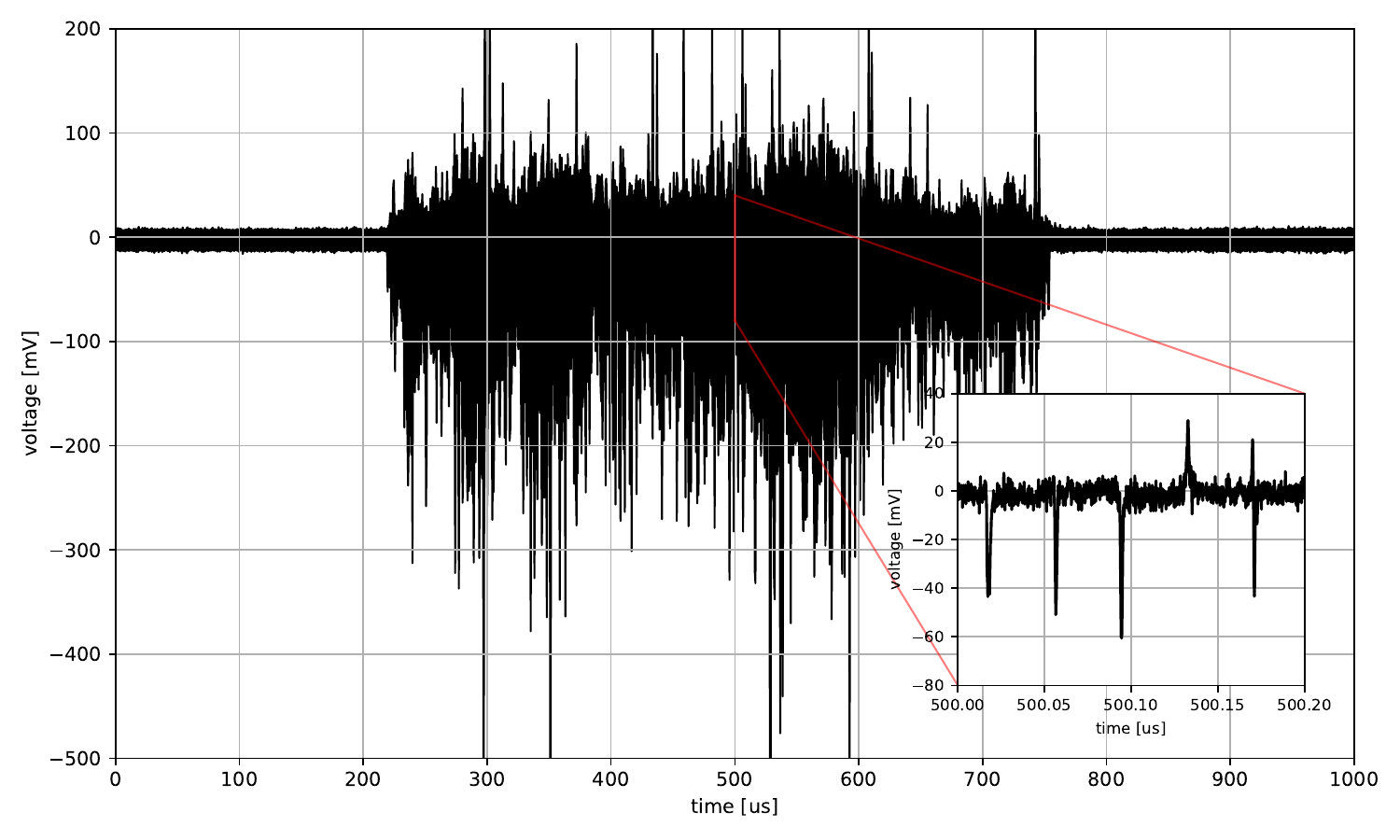}
}
\caption{
A macropulse as recorded by the data acquisition system during a \SI{1}{\milli\second} time window in the 4nA dataset. The inset is a magnified view showing the micropulse structure with a \SI{200}{\nano\second} time window. Signals appear in the negative vertical direction: the pulses in the positive direction are due to electronic induced effects.
}\label{fig:scopetraces}
\end{figure}

Typical data recorded for the 4nA data sample are shown in Fig.~\ref{fig:scopetraces}.
The signals resulting from charge deposits in the silicon pixels result in negative excursions from the baseline with a sharp rising edge and a slightly slower falling edge as equilibrium is restored through the current drawn from the power supply. Already from Fig.~\ref{fig:scopetraces} the features of the cyclotron beam are apparent: the macropulse is seen to be about \SI{540}{\micro\second}, while the micropulses are separated by about \SI{38}{\nano\second}.
Fig. \ref{fig:waveform_typical} shows the waveform corresponding to a typical single particle deposit in the LGAD. It consists of a fast rise time of about \SI{400}{\pico\second} and a slightly longer decay-time of about \SI{600}{\pico\second}.  The width of the signal is less than \SI{2}{\nano\second}.

\subsection{Peak identification}
\label{sec:clusters}

For beam currents up to \SI{4}{\nano\ampere}, the oscilloscope gain was set to \SI{0.125}{\milli\volt\per count} (equivalent to \SI{200}{\milli\volt\per division}), providing a full voltage scale of \SI{2}{\volt} on a \num{10}-division display. For higher beam currents, the gain was doubled. 
With this configuration, the noise level remained at approximately \num{20} ADC counts, independent of the gain setting. This corresponds to a voltage noise of about \SI{2.5}{\milli\volt} for beam currents up to \SI{4}{\nano\ampere} and \SI{5}{\milli\volt} for larger currents. Given that the WavePro 404HD oscilloscope is specified to have a vertical noise floor of \SI{2.06}{\milli\volt_{rms}} at \SI{200}{\milli\volt\per division}, the observed noise is probably dominated by the digitization noise of the oscilloscope.

A simple algorithm was implemented to identify signals from charge deposits in the silicon pixels. For each \SI{1}{\milli\second} time-window (containing \num{1e7} samples at \SI{10}{\giga\mathrm{Samples\per\second}}), a portion of the data was used to characterise the baseline and noise. Specifically, the data acquired during the time between macropulses, just before the \SI{50}{\hertz} cyclotron trigger, was used. This corresponds to the first \SI{150}{\micro\second} or \num{1.5e6} samples (\SI{15}{\percent} of the total). The baseline and noise characterisation was performed individually for each \SI{1}{\milli\second} segment to account for baseline fluctuations over time. The peak detection threshold was set to three times the RMS noise to detect small amplitude peaks.  The exact threshold value was \SI{7.5}{\milli\volt} for beam currents up to \SI{4}{\nano\ampere}, and twice that value for larger beam currents. 
It was required that the signal remains over the threshold for greater than \SI{300}{\pico\second} to reduce noise effects.

For beam currents up to and including
\SI{4}{\nano\ampere}, the peak-finding algorithm is highly efficient. However, the effective doubling of the noise, when the ADC range was increased for the higher beam currents, leads to an inefficiency on single proton detection of 18\%, which was estimated using a portion of the 
\SI{4}{\nano\ampere} data where both ADC settings were used.
The results presented in Sec.~\ref{sec:dosimetry_results} take account of this.

\begin{figure}
    \centering
    \includegraphics[width=0.8\linewidth]{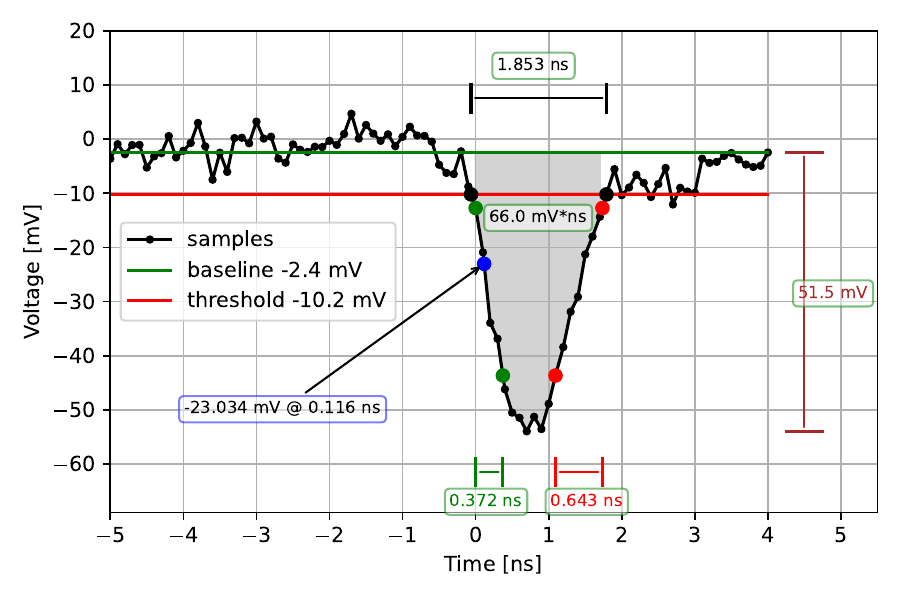}
    \caption{Typical waveform from the lowest intensity dataset (\SI{0.5}{\nano\ampere}), characterized by an amplitude of \SI{51.5}{\milli\volt}, a rise time of \SI{372}{\pico\second}, a fall time of \SI{643}{\pico\second}, and a duration (time over threshold) of \SI{1.853}{\nano\second}. The signal area (\SI{66}{\milli\volt\cdot\nano\second}) between the waveform and the baseline is shaded in light grey.}
    \label{fig:waveform_typical}
\end{figure}

The peak position was determined offline using a Constant Fraction Discriminator (CFD) algorithm to mitigate time-walk effects.  
The position of the particle arrival time was defined as the instant the signal crossed a threshold set at \SI{40}{\percent} of the peak amplitude. Linear interpolation between samples was used to determine the exact crossing time. The signal rise time was defined as the time interval between the signal reaching \SI{20}{\percent} and \SI{80}{\percent} of the amplitude. The signal length (time over threshold) was determined as the time between the last threshold crossing on the rising edge and the first threshold crossing on the falling edge.  Linear interpolation was used to calculate the precise crossing times. The signal area was calculated as the integral of the signal above the baseline, within the time interval in which the signal exceeded the threshold.
An example of the calculation of these quantities is shown in Fig. \ref{fig:waveform_typical}.

At higher beam intensities, multiple protons may traverse a single strip within a micropulse, complicating waveform analysis. If their arrival time differences are shorter than the signal rise-time, the amplitude increases but the peaks cannot be separated,
while if the arrival time differences are longer than the signal rise-time,
distinct peaks may appear but the signal-length and rise-time measurements would become distorted.
The analysis reported later in the paper shows that the micropulse width is only about
\SI{0.5}{\nano\second}, and therefore of the same order as the rise-time shown in Fig.\ref{fig:waveform_typical}.  
Thus, pileup effects do not distort the shapes of the waveform, though they can increase the amplitude.
Nonetheless, the peak-finding algorithm sometimes found additional peaks which were 
attributed to electronic effects including detector ringing and cross-talk.
Therefore, only the first peak was considered by the peak-finding algorithm, and formed the basis for the pileup mitigation strategy described in Section~\ref{sec:poisson_method}.

\subsection{Micropulse periodicity}
\label{sec:micropulse}
The time of arrival, $t$, of a particle (here calculated using the CFD method) can be related to the radio-frequency (RF) of the cyclotron, $f_{\text{RF}}=1/T$, by the following relation:
\begin{equation}
    t = k T + \delta
    \label{eq:deltat}
\end{equation}
where $k$ is an integer and
$\delta$ is an offset ($|\delta$| < T/2).
Note that with the nominal scheme of Fig.~\ref{fig:sketch}, $T\sim \SI{38}{\nano\second}$ while $|\delta|<\sim\SI{1}{\nano\second}$.
Eq.~\ref{eq:deltat} can be used to both measure the RF of the cyclotron and determine the width of the micropulse, by histogramming the values of $\delta$ for different values of $T$, taken to be a free parameter.
When $T$ bears no resemblance to the true cyclotron RF, $\delta$ is a flat distribution between $\pm T/2$, while at the true value, the root-mean-square (RMS) of the $\delta$ distribution is as small as possible with its value giving the width of the micropulse.
In order to suppress ringing effects, especially pronounced at  the highest beam intensities, only the first peak of the micropulse was selected for the histogramming of $\delta$.

\subsection{Particle counting}
\label{sec:poisson_method}

In this work, two algorithmic approaches are used for particle counting and flux estimation. The first method is a straightforward peak counting technique, which identifies and counts individual particle signals based on the peak definition detailed in section \ref{sec:clusters}. The second algorithm assumes a Poisson distribution for particle signal arrivals, as described in this section.

Pileup acts as a contribution to detector inefficiency, since only one signal can be counted per micropulse. However, it does not effect the ability to measure empty micropulses in which no particle traversed the sensor. Assuming that the frequency of particle signals in a given detector channel, within one micropulse, follows a Poisson distribution, the probability that no particle hits a selected channel is
\begin{equation}
    \label{eq:plambda}
    P(0) = e^{-\lambda},
\end{equation}
where $\lambda$ is the average number of signals per micropulse and thus
\begin{equation}
    \label{eq:lambda}
    \lambda=-\ln(f_0),
\end{equation} 
where $f_0$ is the fraction of micropulses with no signal.
Measuring $f_0$ requires choosing a group of $n$ micropulses with $n$ sufficiently large to obtain at least one zero-signal micropulse.  However, $n$ should not be too large since the particle flux can fluctuate significantly within the macropulse, and it is this variation that we are interested in measuring.
The determination of $\lambda$ proceeds by taking all $n$ consecutive micropulses in a recorded macropulse, and for the time coordinate at the centre of the group, calculating $\lambda$ from Eq.~\ref{eq:lambda}.  
If $f_0=0$ for the group, $\lambda$ is assigned the value $-\ln(1/n)$.
The total number of particles in the macropulse is estimated as the sum of the determined $\lambda$s.

The method has been validated using a MonteCarlo sampling procedure that generates particles according to a Poisson distribution and applies this algorithm. 
Tests were performed by varying the generated $\lambda$ in the simulation, mimicking the behaviour of the particle flux observed in the beam test. When testing the algorithm precision in measuring $\lambda$ for multiple values of $n$, it was found that the optimum is affected by the absolute scale of the flux and its variability. As expected, for higher fluxes, larger values of $n$ are required. 
It was found that the optimal values for $n$ in simulation settings that reproduce the flux characteristics of the \SI{1}{\nano\ampere} dataset is $44$, and for nominal currents up to \SI{4}{\nano\ampere} the algorithm performs best with $40<n<100$. 
It was decided to operate the algorithm using $n=44$ for which the simulation of a fully efficient detector predicts a linear response up to $\lambda=3.5$, but, by construction, is not sensitive to fluxes with $\lambda>3.8$. 

\section{Results}
\label{sec:results}

\subsection{LGAD signal characteristics}
\label{sec:clusters_results}

\begin{figure}[!ht]
\centering
\includegraphics[width=0.9\textwidth]{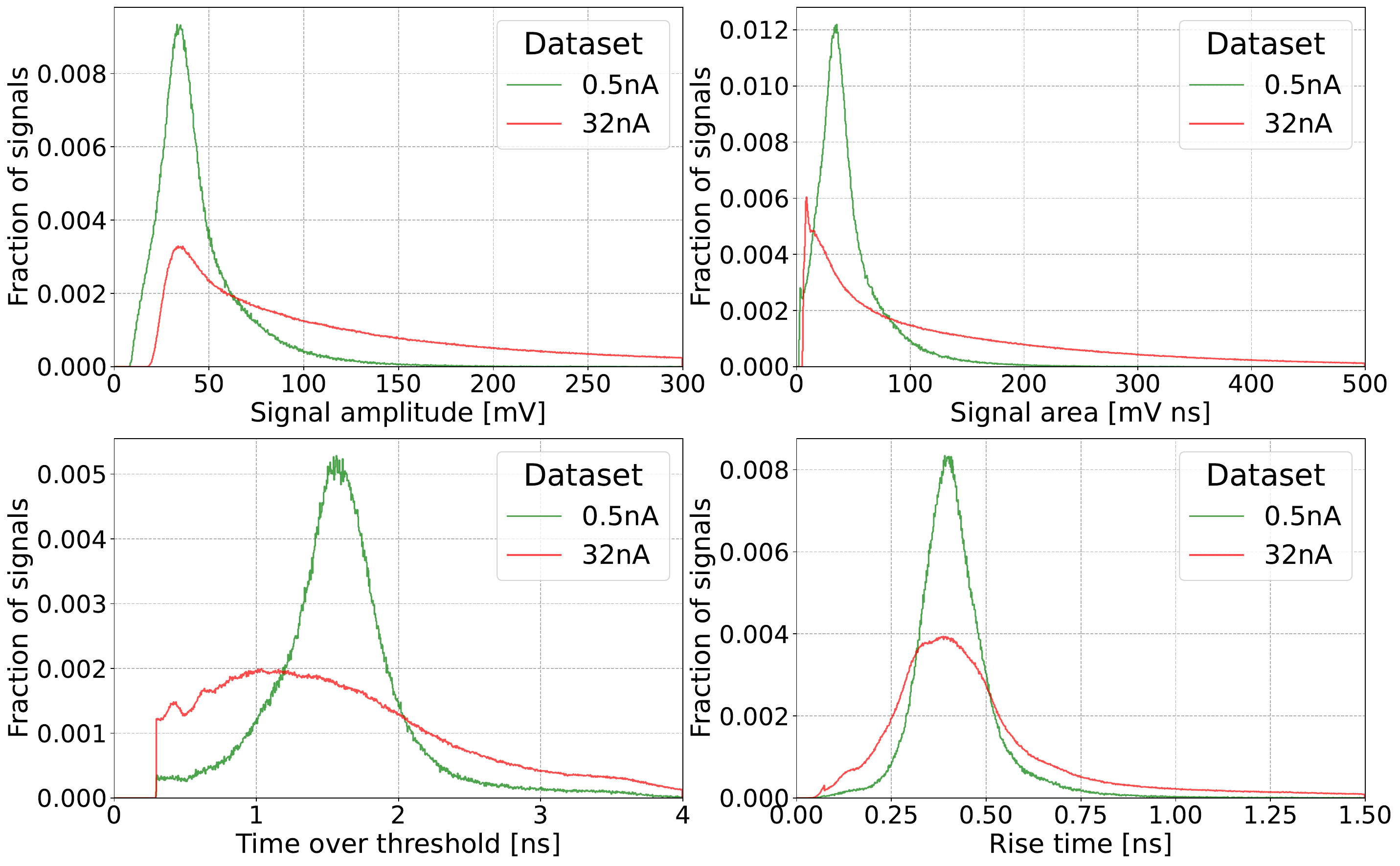}
\caption{
Summary of signal characteristics observed at low beam intensity and high intensity.
All plots are normalised to one.
}\label{fig:pulse_character}
\end{figure}

Typical distributions of the amplitude, area, rise time, and time over threshold (defined in Sec. \ref{sec:clusters}) of selected negative going signals in a detector channel are shown in Fig.~\ref{fig:pulse_character}. These signals were selected using the simple peak counting method described in Sec. \ref{sec:clusters}, for data samples collected at low and high average proton flux values.

The amplitude and area are strongly correlated. The amplitude displays the typical characteristics of a Landau distribution at low proton flux values, while for high fluxes, a more populated right tail forms, most likely driven by pileup events. 
The rise time of the signal has a weak dependence on the charge deposited and is mainly determined by the design of the readout board, whereas the time over threshold shape is heavily altered by overlapping signals caused by pileup.
The rise-time is typically $0.45$\,ns indicating that the intrinsic time precision of the sensor is better than $0.1$\,ns. The total length of the signal is about $1.5$\,ns.

\begin{figure}[t]
\centering
\includegraphics[width=0.7\textwidth]{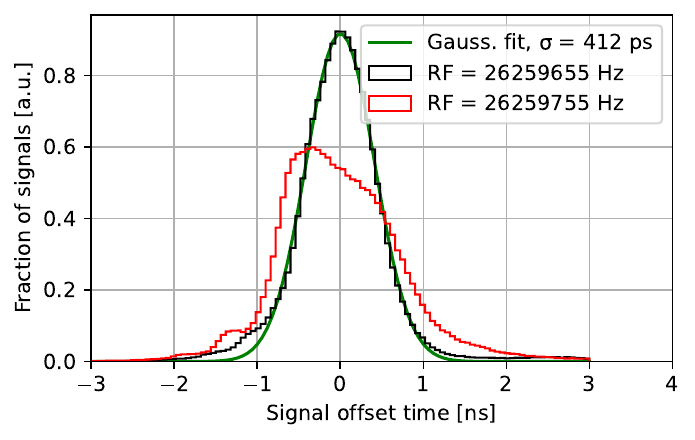}
\includegraphics[width=0.7\textwidth]{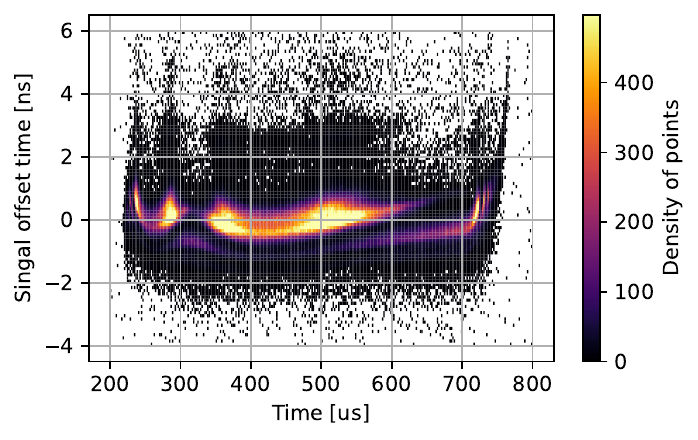}

\caption{Top: Distribution of peak arrival time offsets for a 0.5nA dataset. The black histogram corresponds to the optimal frequency, while in the red histogram the frequency is changed by \SI{100}{\hertz}. The fitted Gaussian function (green) has a width of \SI{412}{\pico\second}.
Bottom: Micropulse arrival times as a function of position in the macropulse.}
\label{fig:rf}
\end{figure}

\subsection{Proton beam dynamics and RF determination}
\label{sec:res_rf}

Fig.~\ref{fig:rf}\ (top) shows the determination of the RF frequency of the cyclotron using the method described in Sec. \ref{sec:micropulse}.
The period is \SI{38.081244}{\nano\second}, which translates into a RF frequency of \SI{26.259646}{\mega\hertz} with a statistical precision well below \SI{1}{\hertz},
consistent with the RF design parameter of \SI{26.26}{\mega\hertz}.
The width of the micropulse is \SI{0.41}{\nano\second}, about half the nominal value indicated in Fig.~\ref{fig:sketch}.
The RF of the machine was found to be stable with a maximum deviation of \SI{10}{\hertz} observed when repeating the analysis on data collected a few hours later.
This is in contrast to the cyclotron investigated in \citep{lgad_heller}, where deviations of about \SI{1}{\nano\second} were found over several minutes.

The average RF frequency serves as a starting point for studies of the proton beam dynamics.
Fig.~\ref{fig:rf}\ (bottom) shows the signal offset time compared to the average for the whole macropulse, as a function of time within the macropulse.
It is seen that the average arrival time of the protons can vary by up to \SI{1}{\nano\second},
a significant amount given that the micropulse width is  
\SI{0.41}{\nano\second}.
The unprecedented timing resolution afforded by the LGAD detectors allows remarkable detail on the operation of the cyclotron to be extracted, and provides a valuable monitoring capability for machine operators. 

\subsection{Evolution of the proton flux within a macropulse}
\label{sec:macropulse}

\begin{figure}
     \centering
     \includegraphics[width=0.7\linewidth]{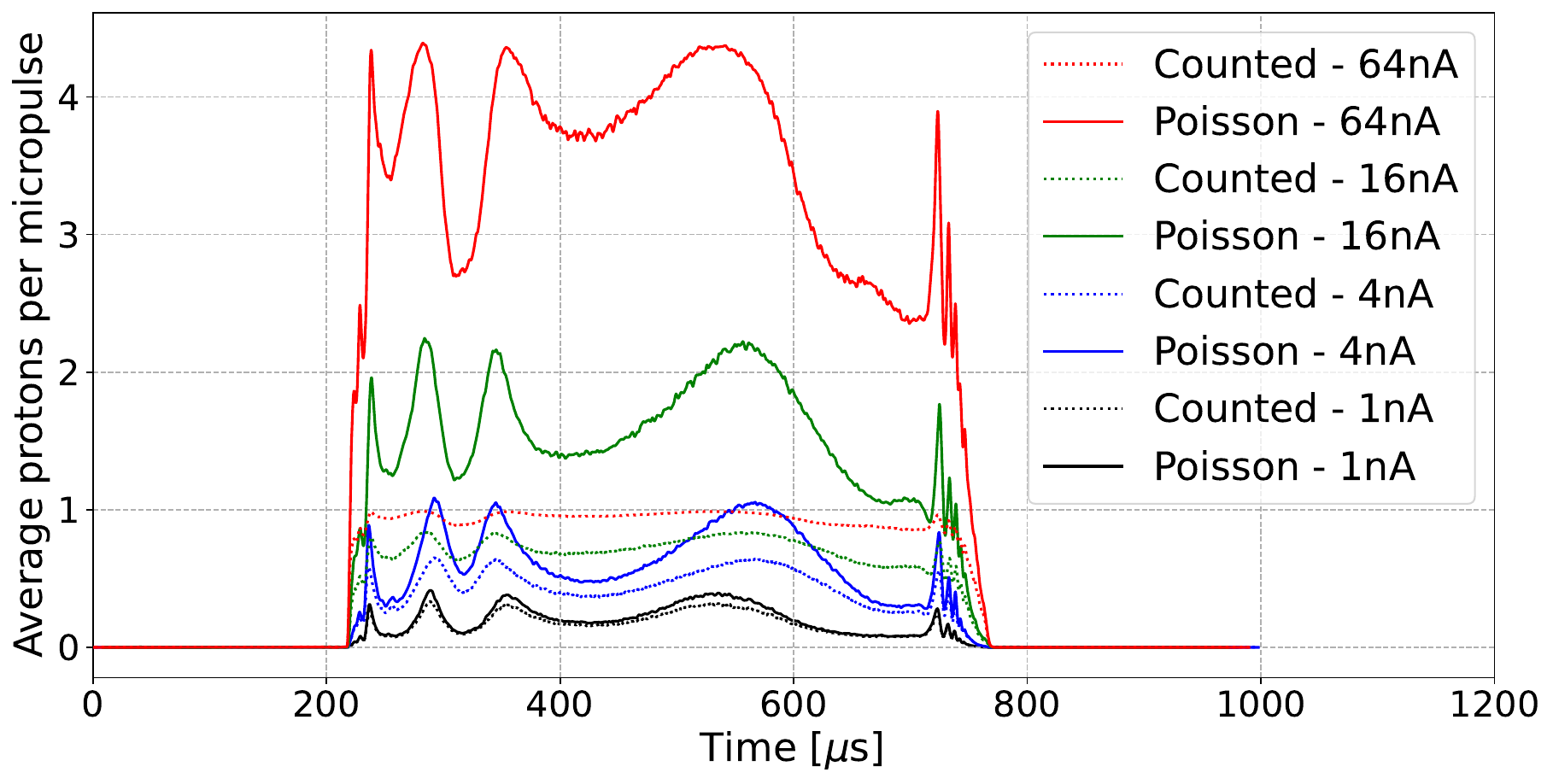}
     \caption{Measured number of protons per micropulse as measured using the simple counting method (dashed lines) and with the Poisson-counting method (solid lines), as a function of time within the macropulse, for four of the data-sets summarised in Table~\ref{tab:design}. 
     }
     \label{fig:macropulse_structure}
 \end{figure}

The variability of the proton flux within the macropulse, as a function of time, was investigated for a wide range of beam intensities.
Two methods were used to estimate the proton flux: simple peak counting, as described in Section \ref{sec:clusters}, and a pileup mitigation method based on Poisson statistics, detailed in Section \ref{sec:poisson_method}. 
 Estimates of the average number of protons per micropulse obtained by simple peak counting and by the Poisson pileup mitigation method are illustrated in Fig.~\ref{fig:macropulse_structure} for four different beam intensities.
At low intensities, both methods give similar results. 
At high intensities, the peak counting method, which by construction can not record more than one proton per micropulse, saturates and underestimates the true number of protons per micropulse.
The Poisson-counting method performs better but, using $n=44$ for the group size, has an upper-bound on the number of measurable protons per micropulse of 3.8, which is hit at certain times for the highest machine intensities. 

The protons flux within the macropulse is not constant and has considerable structure.
A comparison with Fig.~\ref{fig:rf} shows a correlation with the arrival times of the protons and indicates that subtle details of the cyclotron design and operation are being observed for the first time, and are of interest to the beam engineers.

\subsection{Proton Flux Measurement using particle counting}
\label{sec:dosimetry_results}

\begin{figure}[t]
\centering
\includegraphics[width=0.98\textwidth]{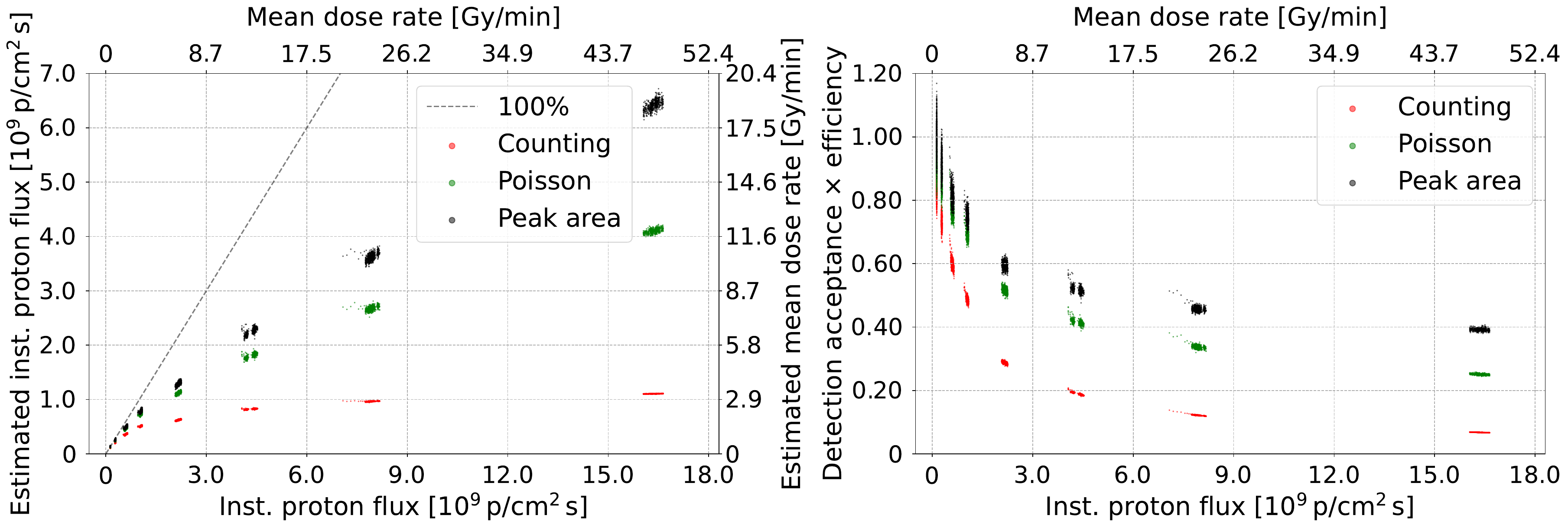}
\caption{Left: Instantaneous flux measured by the LGAD 
compared to the flux measured by the calibrated ionisation chamber.
Results are presented for the simple counting method (red), for the Poisson method (green), and for the total charge deposited (black).  The dashed line indicates a one-to-one correspondence.
Right: Efficiency of the LGAD counting method as a function of instantenous flux.
The efficiency for the deposited-charge measurement is set to 1 at the lowest beam intensity.
}
\label{fig:dosimetry_poisson_and_counting}
\end{figure}

The number of particle crossing the detector determines the instantaneous flux, 
$\Phi_{\rm inst}$ through Eq.~\ref{eq:phi_inst}.
Fig.~\ref{fig:dosimetry_poisson_and_counting} compares this quantity to the flux as measured by the ionisation chamber using Eq.~\ref{eq:flux_ic_inst}.
Results are presented for the different beam intensities summarised in Table~\ref{tab:design},
and for both the simple-counting method
and the Poisson-counting method. 
The figure also provides an axis for the mean
dose rate, in units of \SI{}{\gray\minute}, more commonly used in radiation therapy.

The simple counting method saturates for instantaneous fluxes above \mbox{\SI{3e9}{\per\centi\meter\squared\per\second}} while the Poisson estimation, although not fully
compensating for inefficiencies, 
maintains a growing behaviour, therefore enabling its usage as a monitor of the proton flux, provided a suitable calibration is performed. 

The right panel in Fig.~\ref{fig:dosimetry_poisson_and_counting} translates these results into an efficiency, defined as the ratio between the LGAD and the ionisation chamber measurements.
As long as the flux remains below approximately 3.5 particles per micropulse, preventing saturation of the Poisson method, Fig.\ref{fig:dosimetry_poisson_and_counting} can be utilised to calibrate the LGAD response. This calibration would allow for the conversion between the LGAD-estimated dose rate and the ion chamber mean dose rate. Consequently, the LGAD detector shows considerable promise as a beam monitor due to its sub-$\SI{1}{\micro\second}$ temporal response, making it an excellent candidate for time-critical applications.

For fluxes above $\lambda=3.5$ particles per micropulse, corresponding to about \SI{7.5}{\gray\per\second} (see Table~\ref{tab:design}), the rates are too high to get an accurate response.
An attempt was made to improve the sensitivity at high fluxes by increasing the group-size, $n$, in the Poisson algorithm from 44 to 1000, but this led to  minimal improvements,  due to the presence of spurious positive pulses and ringing effects of the electronics at high proton fluences, which sometimes cancel genuine (negative) signals, leading to false zeroes in the Poisson algorithm.
Some modifications to the sensor and electronics will be necessary to mitigate these effects at the highest beam intensities, as discussed in Sec.~\ref{sec:discussion}.

In the context of proton therapy, sensitivity to higher beam intensities is desirable. Beam currents of 20 nA were used for treatment of eye tumours in the AIC-144 cyclotron, which corresponds to roughly 5 particles per micropulse in a single strip of the LGAD sensor. Therefore, to apply this detector in clinical settings, a smaller strip or pad size would be necessary.

We also note that even at the lowest intensities where pileup-induced inefficiency contributions are negligible,
the efficiency does not reach $100\%$.
LGADs are known to have efficiencies close to $100\%$, however the active area is generally smaller than the pad size that can be measured with a microscope. In Villareal et al. \citep{MARTIVILLARREAL2023167622}, LGAD sensors (type “E”, \SI{216}{\micro\meter} pitch) were characterized using a laser probe, yielding an inter-strip dead region of \SI{80}{\micro\meter}—22\% larger than the nominal \SI{66}{\micro\meter}. This increase is attributed to incomplete charge collection in the transition between the gain layer periphery and the adjacent JTE implant \citep{paternoster2021novel}. Assuming an \SI{80}{\micro\meter} dead region for type “T” sensors, the effective sensitive area is reduced by about 6\%, which may explain the reduced efficiency even at the lowest fluxes.  

\subsection{Proton Flux Measurement using charge deposits}

The charge deposited in the LGAD sensor is related to the integrated signal area, defined as the sum of the area under all signals in a macropulse (See Section~\ref{sec:clusters} and the top-right panel of Fig.~\ref{fig:pulse_character}). 
This quantity is correlated to the number of particles crossing the detector since more particles deposit more charge.
Therefore, as an alternative to particle counting (or zero-counting in the Poisson method), we summed the signal area for all micropulses in a macropulse and plotted this quantity against the instantaneous flux measured by the ion chamber.
The results are also presented in Fig.~\ref{fig:dosimetry_poisson_and_counting}.
For the efficiency calculation,
the factor relating the total charge and the dose is taken from the lowest intensity beam data, so that by definition, the efficiency here is 1.
Compared to the particle-counting results, the total charge appears to have a more linear response with the beam intensity, and therefore the summed-charge deposited in the LGAD presents an alternative observable to particle-counting for monitoring the beam.

\section{Discussion}
\label{sec:discussion}

The results illustrated in this work show that LGADs represent a promising solution both for beam monitoring and dosimetry at high proton fluxes. 
Their short signals allow the structure of the beam, consisting of micropulses and micropulses, to be resolved with unprecedented precision. Furthermore, features that would be completely undetectable with standard instrumentation can be viewed.
In particular, we note variations in the intensities of the micropulses with time, and can also discern temporal variations in the arrival time of each micropulse.

The potential for use as a beam monitor has been demonstrated, either by counting individual protons passing through the device, or by integrating the charge deposited.
The response scales with the beam intensity as measured by ionisation chambers, although it is not precisely linear, thus requiring some calibration if the devices were to be used in a medical environment.

The specific advantage of LGADs lies in their time resolution:  we have shown that signals can be measured within a time window of \SI{2}{\nano\second}, and doses can be estimated from few-microsecond long measurements.
Such precision is probably only of academic interest, but the ability to
measure the delivered dose over single cyclotron pulses of \SI{540}{\micro\second} has practical applications in FLASH therapy or in time-critical scenarios, and is orders of magnitude better than the temporal resolution of industry-standard ion chambers. 

Nevertheless, for FLASH applications, the sensor efficiency at high dose rates needs to be improved. Even though a portion of the inefficiency might be due to an inaccurate estimation of sensor active area \citep{data2024novel}, most of the efficiency loss observed is driven by effects related to overlapping signals from multiple particles in pileup events. 
Induced positive peaks and electronic ringing complicate the peak detection algorithm and affect particle counting. In addition,
the voltage baseline was observed to drift up to about \SI{50}{\milli\volt} at the largest beam intensities, during the recovery from very high excursions.

While the full mitigation of these effects will require further effort, certain straightforward improvements could quickly be implemented. The usage of sensors with smaller active area would suppress pileup effects, thereby eliminating the primary source of inefficiency: for example, replacing the strips with \SI{200}{\micro\meter}$\times$\SI{200}{\micro\meter} pixels would allow particle-counting at \SI{400}{\gray\per\second}.
A further revision and optimization of the readout electronics could suppress ringing effects and positive excursions, while the peak detection algorithm could be improved to correct for voltage baseline drifts and to identify multiple peaks within a single micropulse. 

To deploy this prototype setup as a real-world device, the DAQ system would require substantial improvements to remove the reliance on an oscilloscope and to provide an instantaneous dose measurement.
High acquisition efficiency is crucial for tracking the evolution of the macropulse structure over extended periods, as beam tuning can significantly alter it. A dedicated DAQ system based on an FPGA and a PC with substantial RAM, as reported in \citep{hueso2022dead}, could achieve 100\% efficiency in processing waveforms for our setup while maintaining the required response time.

\section{Conclusions}
\label{sec:conclusions}

This study demonstrated the potential of LGAD sensors as beam monitors in proton therapy, providing a precise characterisation of the macropulse and micropulse structures of a cyclotron beam. The subnanosecond timing resolution of LGADs enabled detailed observation of the beam structure, surpassing the capabilities of conventional ionisation chambers. 

Pileup effects were found to manifest themselves primarily as increased signal amplitude rather than distinct peak separation,
and algorithms aimed at reducing these effects, relying on zero-counting or the measurement of the peak areas, were developed.
An estimation of the proton flux was performed with multiple techniques and the LGAD results were compared with ionisation chamber measurements. 
Accurate flux measurement were achieved up to instantaneous dose rates of about \SI{7.5}{\gray\per\second}.

Future studies will focus on optimising signal processing techniques, reducing noise, and exploring alternative sensor geometries to improve detection efficiency. LGADs, with further refinement, could become a viable alternative for high-precision proton beam monitoring in clinical and research environments.

\roles{The authors contributed to this work according to the CRediT (Contributor Roles Taxonomy) methodology as follows: 
Conceptualization – A.B., L.G., R.M., N.M., J.S.; 
Methodology – L.G., K.M., M.N., P.R.; 
Software – A.B., L.G., R.M., N.M., K.M., M.N., P.R.;
Validation – A.B., L.G., N.M., R.M., T.N.;
Formal Analysis – A.B., L.G., N.M., R.M.;
Investigation – A.B., L.G., R.M., N.M., T.N., J.S. T.S.;
Data Curation – L.G., K.M., M.N., P.R.;
Writing – Original Draft – A.B., L.G., R.M.; 
Writing – Review \& Editing – A.B., L.G., R.M., N.M., K.M., M.N., P.R., J.S., T.S.; 
Visualization – A.B., L.G., R.M.; 
Supervision – L.G., R.M., N.M.; 
Project Administration – R.M., L.G., J.S.; 
Funding Acquisition – L.G., R.M., J.S.

All authors have read and agreed to the published version of the manuscript.
}

\ack{
We gratefully acknowledge the INFN group of Torino and FBK for the outstanding work in designing and manufacturing the silicon sensors, which was funded by the INFN-CSN5 experiment MoVeIT. The authors sincerely thank the AIC-144 cyclotron operators and especially the Department of Radiation Research and Proton Radiotherapy crew for their vital support and professionalism during the experimental campaign.

The authors used ChatGPT (OpenAI) and Gemini (Google) for text editing and language refinement. All final revisions were reviewed and approved by the authors, who assume full responsibility for the content.
}

\funding{
This project has received funding from the European Union’s Horizon Europe Research
and Innovation programme under Grant Agreement No 101057511 (EURO-LABS-ALTO-001) for the irradiation of the LGAD sensors.  Andrea Bellora received support from the Excellence Initiative - Research University (IDUB) at AGH University in Krakow. The research presented in this article was supported by the funds allocated to the AGH University of Science and Technology by the Polish Ministry of Science and Higher Education. The authors gratefully acknowledge the Polish high-performance computing infrastructure PLGrid (HPC Centre: ACK Cyfronet AGH) for providing computer facilities and support within computational grant no. PLG/2024/017365. 
}

\bibliography{Bibliography}

@techreport{Albrow:2014lrm,
    author = "{The CMS and TOTEM Collaborations}",
    collaboration = "CMS, TOTEM",
    title = "{CMS-TOTEM Precision Proton Spectrometer}",
    number = "CERN-LHCC-2014-021, TOTEM-TDR-003, CMS-TDR-13",
    year = "2014",
    url = "http://cds.cern.ch/record/1753795",
    institution = "CERN"
}

@article{Isidori:2021cwt,
    author = "Isidori, Tommaso and others",
    title = "{Performance of a low gain avalanche detector in a medical linac and characterisation of the beam profile}",
    eprint = "2101.07134",
    archivePrefix = "arXiv",
    primaryClass = "physics.ins-det",
    doi = "10.1088/1361-6560/ac0587",
    journal = "Phys. Med. Biol.",
    volume = "66",
    number = "13",
    pages = "135002",
    year = "2021"
}

@article{swakon2010facility,
  title={Facility for proton radiotherapy of eye cancer at {IFJ PAN} in {Krakow}},
  author={Swakon, J and Adamczyk, D and Dulny, B and others},
  journal={Radiation Measurements},
  volume={45},
  number={10},
  pages={1469--1471},
  year={2010},
  publisher={Elsevier}
}

@article{rydygier2018radiotherapy,
  title={Radiotherapy proton beam profilometry with {scCVD} diamond detector in single particle mode},
  author={Rydygier, Marzena and others},
  journal={Radiation protection dosimetry},
  volume={180},
  number={1-4},
  pages={282--285},
  year={2018},
  publisher={Oxford University Press}
}

@article{hueso2022dead,
  title={A dead-time-free data acquisition system for prompt gamma-ray measurements during proton therapy treatments},
  author={Hueso-Gonz{\'a}lez, Fernando and others},
  journal={Nuclear Instruments and Methods in Physics Research Section A: Accelerators, Spectrometers, Detectors and Associated Equipment},
  volume={1033},
  pages={166701},
  year={2022},
  publisher={Elsevier}
}

@article{data2024novel,
  title={A novel detector for {4D} tracking in particle therapy},
  author={Data, EM  and others},
  journal={Nuclear Instruments and Methods in Physics Research Section A: Accelerators, Spectrometers, Detectors and Associated Equipment},
  volume={1068},
  pages={169690},
  year={2024},
  publisher={Elsevier}
}

@book{paganetti2025proton,
  title={Proton therapy physics},
  author={Paganetti, Harald},
  year={2025},
  publisher={CRC press}
}

@article{schippers2009beam,
  title={Beam delivery systems for particle radiation therapy: Current status and recent developments},
  author={Schippers, JM},
  journal={Reviews of Accelerator Science and Technology},
  volume={2},
  number={01},
  pages={179--200},
  year={2009},
  publisher={World Scientific}
}

@misc{newhauser2009international,
  title={International commission on radiation units and measurements report 78: Prescribing, recording and reporting proton-beam therapy},
  author={Newhauser, Wayne},
  year={2009},
  publisher={Oxford University Press}
}

@article{ziegler2010srim,
  title={{SRIM}--The stopping and range of ions in matter (2010)},
  author={Ziegler, James F and Ziegler, Matthias D and Biersack, Jochen P},
  journal={Nuclear Instruments and Methods in Physics Research Section B: Beam Interactions with Materials and Atoms},
  volume={268},
  number={11-12},
  pages={1818--1823},
  year={2010},
  publisher={Elsevier}
}

@phdthesis{shakarami2021development,
  author       = {Shakarami, Zahra},
  title        = {Development of a Novel Solid State Detector for Beam Monitoring in Proton Therapy},
  school       = {Università degli Studi di Torino},
  year         = {2021},
  address      = {Torino, Italy},
  type         = {PhD Thesis},
}

@phdthesis{abujami2024development,
  author       = {Abujami, Mohammed},
  title        = {Development and Test of an Innovative Particle Counter for Beam Monitoring in Particle Therapy},
  school       = {Università degli Studi di Torino},
  year         = {2024},
  address      = {Turin, Italy},
  type         = {PhD Thesis}
}

@article{monaco2023performance,
  title={Performance of {LGAD} strip detectors for particle counting of therapeutic proton beams},
  author={Monaco, Vincenzo and others},
  journal={Physics in Medicine \& Biology},
  volume={68},
  number={23},
  pages={235009},
  year={2023},
  publisher={IOP Publishing}
}

@article{paternoster2021novel,
  title={Novel strategies for fine-segmented low gain avalanche diodes},
  author={Paternoster, G and others},
  journal={Nuclear Instruments and Methods in Physics Research Section A: Accelerators, Spectrometers, Detectors and Associated Equipment},
  volume={987},
  pages={164840},
  year={2021},
  publisher={Elsevier}
}

@article{pellegrini2014technology,
  title={Technology developments and first measurements of Low Gain Avalanche Detectors ({LGAD}) for high energy physics applications},
  author={Pellegrini, Giulio and others},
  journal={Nuclear Instruments and Methods in Physics Research Section A: Accelerators, Spectrometers, Detectors and Associated Equipment},
  volume={765},
  pages={12--16},
  year={2014},
  publisher={Elsevier}
}

@article{li2012beyond,
  title={Beyond Gaussians: a study of single-spot modeling for scanning proton dose calculation},
  author={Li, Yupeng and others},
  journal={Physics in Medicine \& Biology},
  volume={57},
  number={4},
  pages={983},
  year={2012},
  publisher={IOP Publishing}
}

@article{lgad_heller,
title={Demonstration of {LGADs} and {Cherenkov} gamma detectors for prompt gamma timing proton therapy range verification},
author={Heller, R and others},
journal={IEEE Trans Radiat Plasma Med Sci.},
volume={9},
number={4},
pages={508},
year={2024}
}

@article{MARTIVILLARREAL2023167622,
title = {Characterization of thin {LGAD} sensors designed for beam monitoring in proton therapy},
journal = {Nuclear Instruments and Methods in Physics Research Section A: Accelerators, Spectrometers, Detectors and Associated Equipment},
volume = {1046},
pages = {167622},
year = {2023},
issn = {0168-9002},
doi = {https://doi.org/10.1016/j.nima.2022.167622},
url = {https://www.sciencedirect.com/science/article/pii/S0168900222009147},
author = {O.A. {Marti Villarreal} and others}
}

\end{document}